# ANAP: Anonymous Authentication Protocol in Mobile Ad hoc Networks

Tomasz Ciszkowski  
*Email: T.Ciszkowski@tele.pw.edu.pl*

Zbigniew Kotulski  
*Email: Z.Kotulski@tele.pw.edu.pl*

Warsaw University of Technology

**Abstract**

The pervasiveness of wireless communication recently gave mobile ad hoc networks (MANET) a significant researcher's attention, due to its innate capabilities of instant communication in many time and mission critical applications. However, its natural advantages of networking in civilian and military environments make them vulnerable to security threats. Support for an anonymity in MANET is an orthogonal to security critical challenge we faced in this paper. We propose a new anonymous authentication protocol for mobile ad hoc networks enhanced with a distributed reputation system. The main its objective is to provide mechanisms concealing a real identity of communicating nodes with an ability of resist to known attacks. The distributed reputation system is incorporated for a trust management and malicious behavior detection in the network. The end-to-end anonymous authentication is conducted in three-pass handshake based on an asymmetric and symmetric key cryptography. After successfully finished authentication phase secure and multiple anonymous data channels are established. The anonymity is guaranteed by randomly chosen pseudonyms owned by a user. Nodes of the network are publicly identified and are independent of users' pseudonyms. In this paper we presented an example of the protocol implementation.

## 1 Introduction

The wireless and mobile communication is a incessantly evolving area providing an ability of flexible and convenient communication with phones, mobile or handheld computers, and almost every contemporary electronic home devices. For known wireless technologies such as Wi-Fi (802.11), IrDA, Bluetooth, communication between network devices is usually realized in a point-to-point model. In mobile ad hoc networks (MANET) devices, interchangeably called nodes, do not have a priori knowledge of network topology surrounding them and they have to discover it. The communication in stuctureless environment assumes a model of intermediate nodes which interconnect devices lying beyond its radio range. In order to make that multihop network operational the self-configuring and self-organizing mechanisms must be introduced to the routing protocols. The basic idea of routing protocol for ad hoc network is to announce a new node presence and listen to broadcast announcements from its neighbors. The way how the node learns about new neighboring nodes is based on the path discovery algorithms and characterizes particular routing protocol. The are several different approaches delivering discovery path feature in ad hoc networks however two groups of them are most popular: reactive (on-demand) and proactive (table-driven). The first type of routing addresses destination node path resolving on demand of source node whereas the second approach is more preventive and continuously is trying to keep routing tables up to date by monitoring the nearest neighborhood. The comparison and features description routing protocols for ad hoc networks may be found in [7], [8], and [17] and is beyond of this paper.

In recent years mobile ad hoc networks have received significant researcher's attention due to capabilities of establishing an instant communication in many time-critical and mission-critical applications. Many security protocols have been devised to protect a communication in ad hoc networks [10], [20], however the only few of them address privacy guarantees [3], [4], and [5]. Leaving mobile nodes traceable by wireless traffic and data analysis makes the anonymity support in MANET critical challenge.

In this paper we propose a new anonymous authentication protocol for mobile ad hoc networks enhanced with a distributed reputation system. The main objective of this work is to provide protocol with mechanisms concealing real identity communicating nodes and resisting to known attacks [6], [23]. The distributed reputation system is incorporated in order to build and manage trust of communicating nodes. The trust knowledge reflects trustworthy and malicious activity in the network effectively supporting anonymous authentication and path discovery phases. Proposed protocol delivers secure exchange data links based on on-demand routing approach [1], [7], and [8].

The following sections present related work and cover in details a protocol construction supplemented by its anonymous properties analysis. The last section presents some concluding remarks and further research directions.

## 2 Background

A set of informal notions of anonymity and its related properties can be found in [23] and [6] where Pfitzman et al. characterize general anonymous system with identity management regarding fixed network topology easily applicable for MANET. There was several work done in a scope of securing mobile ad hoc networks [10], [20], much less [15] addressing privacy aspects.

In terms of authentication in mobile ad hoc networks there are three conventional approaches and they derive from variety of existing protocols for mobile ad hoc networks [10], [20].

- First method is based on shared session key widely distributed in the network [18]. This approach does not provide anonymity and is vulnerable to single node compromise.
- Second one assumes sharing pairwise keys between all nodes in the networks but it suffers from lack of scalability. In this case for N nodes $N(N-1)/2$ keys are required a priory to allow communication [8].
- The third approach takes an advantage of scalable public key cryptography and digital certificates providing network nodes with mutually authentication with challenge-response mechanism [9], [19]. However this method is flexible and efficient the trusted authority is required still.

All of aforementioned methods assume existence of trusted TA or certificate authority CA dealing with key setup phase. Online and distributed CA providing a keys management in anonymous environment are not trivial task and pay researchers' attention [11],[10],[20].

Anonymity schemes in mobile ad hoc networks were proposed in [3], [4], and [5]. ANODR [5] is based on on-demand with identity free routing protocol using a symmetric cryptography with a 'trapdoor boomerang onion' (TBO) approach, similar to onion routing [24] used by Chaum in [23]. The trapdoor mechanism consists of sending cryptographically secured message which may be opened only by intended party. ANODR author pointed out in [15] its low performance in highly mobile networks. In MASK [3] protocol proactive and reactive approach are applied simultaneously. A priori anonymous links are established with all neighboring nodes using a symmetric cryptography and trusted authority. The path discovery process is conducted in an on-demand manner and mutually authenticated nodes participate in the end-to-end communication. Already established path may consist of several multipath channels however the source and destination nodes become unauthenticated. In SDAR [4] communication between source and destination is based on a public key cryptography. Additionally the destination node shares a symmetric session key with each intermediate nodes and uses them to secure discovery path process. This protocol take an advantage of onion and on demand routing. Messages in SDAR are large and strongly depend on the number of hops. Nevertheless, SDAR as first anonymous protocol for mobile ad hoc networks introducing a trust management system. However, this system supports only three levels of permissible reputation limiting therefore its efficiency.

The main objective of anonymous communication in ad hoc network is to provide privacy for all of its users represented by nodes. Demanding for privacy guaranties imposes series of requirement for protocol construction and addresses problem of many known attacks. In closed networks the level of trust is usually managed by authentication mechanisms. In anonymous and mobile ad hoc networks the authentication does not completely solve problems related to nodes misbehavior. Additionally, the anonymity of users creates an environment where personal and direct incentives to be well cooperative are difficult to meet. Therefore in our proposal of the anonymous authentication protocol we employ an enhanced reputation system to detect and avoid cooperation of hostile nodes.

In terms of anonymous communication in mobile ad hoc networks the misbehaving means deviation from basic and regular functions such as routing (next hop finding) and forwarding (relaying a packets). The trust and reputation definitions, which can be found in [22], determine reputation as a level of trust we can assign to every node taking into account a probability of future intentions and behaviors. The trust is often treated as a measure of reputation, but sometimes they are synonyms. A discussion of different approaches in providing reputation can be found in [14]. All of them incorporate the monitoring system as a principle module of misbehavior detection [22] which was effectively improved by exchanging the second-hand information [16].

Liu in [14] presented an enhanced model of reputation for mobile ad hoc networks. This work takes into consideration self experience of nodes, time and context dependency and introduce the definitions of services and recommendation reputation. In [4] for anonymous MANET a three level community management trust was introduced, in which every node is a central node of the community consisting of the rest nearest one-hop neighbors. Reputation is created locally, based on detecting packet dropping and modifications. It determines one of three general classes in all network which is assigned to the node. Based on class membership route discovery is conducted. The main drawback of this approach is low granularity of class and not using a second-hand information.

Our approach for anonymous mobile ad hoc networks is a simple adaptation of Liu [14] reputation model. We propose our own method of evaluating recommendation reputation considering past experience and recommendation reputation of voters (recommenders).

## 3 Anonymous authentication protocol in ad hoc networks

The proposed protocol consists of two complementary modules. First is in charge of anonymous authentication, establishing and maintaining bidirectional routes form source to the destination whereas second one monitors activity of nodes being along the existing or discovering path and evaluate their reputation.

The process of mutual and anonymous authentication is based on the public-key cryptography and is conducted in three-pass handshake. First phase is aimed to initialize a discovery path from source **S** to destination **D** node. Node **S** and **D** know their public-keys with related pseudonyms which are distributed during network setup by trusted authority **TA**. Device **S** broadcasts a request of discovery path to node **D** without revealing its identity. Every node is obligated to relay this message through the entire network, in particular delivering it to the destination **D**. In the same way every node saves the reverse path form **D** to **S**. Node **D** verifies the validity of received message and if succeeds in second phase prepares and unicastly sends response to source node **S** using already created reverse path. The reply message travels hop-by-hop to the node **S**. When this message gets to the legitimate node **S** then its authenticity is verified. Successfully finished two phases determine bidirectional path from **S** to **D**. Generally, it is acceptable to establish several paths between nodes **S** and **D** (e.g. in order to increase performance of network), however initializing node **S** may choose only a subset of offered by **D** paths taking into account its preferences according to reputation system. This is done in former, confirmation phase of handshake. Supplementary, node **D** generates key material for session keys, one for every path, used by symmetric cryptography algorithms, thereby securing established data path.

The above authentication process is described in details in the further subsections. To clarify and transparently present every cryptographic and protocol operations we made following assumptions and notions (Table 1):

- There is, outside of the network, a trusted authority **TA** managing pseudonyms and related to them pairs of public and private keys.
- A couple of nodes willing to communicate know the pair <pseudonym, public key> of each other.
- Radio links are bidirectional giving an ability to reputation system of interested node to monitor traffic in close node's neighborhood. It is required by nodes to support on link layer level promiscuous mode capabilities.
- Nodes have enough computation power to perform all introduced cryptographic operations.
- In the ad hoc network there are nodes with malicious intentions.

| | |
|---|---|
| **h** | hash function |
| $E_{PK\ X} / D_{SK\ X}$ | asymmetric encryption/decryption with **X's** public/private-key $PK_X /SK_X$ |
| $E_{K\ Y} / D_{K\ Y}$ | symmetric encryption/decryption with **Y's** session key $K_Y$ |
| $M_Z$ | general purpose **Z's** message - data payload |
| $Sig(M_Z)$ | signature of message $M_Z$ |
| $SecK(K_Y)$ | protocol for secure key $K_Y$ transport, (e.g. Shamir's no key protocol, Elgamal) |
| **X** | node identifier (e.g. Medium Access Control address of IEEE 802.11) |
| $X_{ID}$ | pseudonym of user who owns **X** node |
| $R_{ID}$ | requested path identifier equals **h( destination $X_{ID}$ )**, reverse path ID |
| $F_{ID}$ | forward path ID |
| **T** | type of messages **REQ, REP, ACK, ERR,** request, reply, acknowledge respectively |
| **Seq** | request path sequence number |
| **ISeq** | internal sequence number delivering randomization and correct packet reassembling |
| **P** | padding of length $L_P$ |
| $L_P$ | padding length |
| $RT_X$ | **X's** reverse table of format <$R_{ID}$, **Seq**, **REQ Source Node**> |
| $RqT_X$ | **X's** request table containing all sent requests **REQ** for anonymous authentication; its format is <$R_{ID}$, **Seq, Reply**>, where **Reply** is a binary value and indicates a state whether response to this request was received |
| $FT_X$ | **X's** forward table with format <$R_{ID}/F_{ID}$, **Seq, From node, $K_{FROM}$, To node, $K_{TO}$**> |
| **TIMER** | clock determining idle time after which a particular entry in **RT** should be removed |

**Table 1. Notions and their definitions applied in protocol operations**

A mobile ad hoc network is initialized with an outside trusted authority **TA**. The **TA** delivers to all users possessing mobile nodes ability of anonymous participation and communication in their network. The issue of online trusted authority was discussed in [11], [21] and is on going authors' research direction.

Every user generates list of its pseudonyms **<X1, X2, X3,…,XN>** and for each of them pair of private and public key **<SK,PK>$_X$** respectively. The **TA** has its own pair of **<SK,PK>$_{TA}$** used for binding pseudonyms and public key **PK$_X$** with user's real identity **UI**. This step is conducted during preliminary setup phase and it allows creating revocable anonymity. The **TA** receives message **M$_{UI}$** encrypted with its public key, decrypt it and stores bindings of format **<UI, PK$_{XN}$, Xn, Sig$_{TA}$(PK$_{XN}$, Xn)>** in a confidential list **CL**. In the second phase **TA** securely distribute a public list **PL** of all network participants concealing real identity **UI**. For two users **UI$_X$, UI$_Y$** described process is illustrated in Figure 1.

| Step | Sender | Action | Receiver |
|------|--------|--------|----------|
| 1 | UI$_X$: M$_{UIX}$ = <X1, PK$_{X1}$>,<X2, PK$_{X2}$> | | |
| 2 | UI: E$_{PK\_TA}$(M$_{UI\_X}$) | → | TA: CL← <UI$_X$, PK$_{X1}$, X1, Sig$_{TA}$(PK$_{X1}$, X1)> |
|   |   |   | CL← <UI$_X$, PK$_{X2}$, X2, Sig$_{TA}$(PK$_{X2}$, X2)> |
| 3 | UI$_Y$: M$_{UIY}$ = <Y1, PK$_{Y1}$>,<Y2, PK$_{Y2}$> |   | TA: PL← <PK$_{X1}$, X1, Sig$_{TA}$(PK$_{X1}$, X1)> |
|   |   |   | PL← <PK$_{X2}$, X2, Sig$_{TA}$(PK$_{X2}$, X2)> |
| 4 | UI$_Y$: E$_{PK\_TA}$(M$_{UI\_Y}$) | → | TA: CL← <UI$_Y$, PK$_{Y1}$, Y1, Sig$_{TA}$(PK$_{Y1}$, Y1)> |
|   |   |   | CL← <UI$_Y$, PK$_{Y2}$, Y2, Sig$_{TA}$(PK$_{Y2}$, Y2)> |
| 5 |   |   | TA: PL← <PK$_{Y1}$, Y1, Sig$_{TA}$(PK$_{Y1}$, Y1)> |
|   |   |   | PL← <PK$_{Y2}$, Y2, Sig$_{TA}$(PK$_{Y2}$, Y2)> |
| 6 | TA: E$_{PK\_X1}$(PL) | → | UI$_X$: D$_{SK\_X1}$(PL) |
| 7 | TA: E$_{PK\_Y1}$(PL) | → | UI$_Y$: D$_{SK\_Y1}$(PL) |

**Figure 1. Network setup phase. Pseudonyms generation and keys issuing.**

The following description is based on simplified ad hoc network model composed of five nodes **S, A, B, C, D** illustrated in Figure 2. Source and destination nodes are denoted by **S** and **D** respectively and the rest group of nodes intermediate in communication between them.

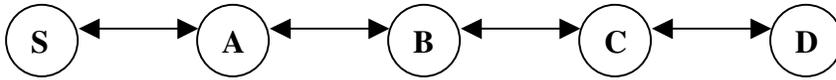

**Figure 2. A simplified ad hoc network model. Five nodes actively participate in anonymous data path establishment. The source node S exchange data with the node D.**

### *3.1 Anonymous authentication initialization – Phase I*

The first phase initializes authentication and path discovery processes. All steps related to this part of protocol are illustrated in Figure 3 with notions reflecting to the example depicted in Figure 2. Similar to on-demand routing approach, the source node **S** generates anonymous request message **REQ** and broadcasts it to the network. The message **M$_S$** contains the pseudonym of a user who owns node **S**, for simplicity denoted as node's pseudonym **S$_{ID}$**, indicating which public-key should be used by **D** in response message. Internal sequence number **ISeq** is used during packet reassembling process and in conjunction with padding is used as a randomizer of the message. The path identifier **R$_{ID}$** is a hash value of destination node's pseudonym and reveals temporary the destiny of request message. This parameter points out that only real destiny **D** is able to decrypt **REQ** message. A sequence number **Seq** of request packet determine the validity of the route identified by **R$_{ID}$**. This is a know mechanism from on-demand routing protocols such as AODV [26]**,** DSR [12] which provide routing optimization features described briefly later. As soon as source node prepares and sends the request message it saves in its request table **RqT$_S$** pair of values **<R$_{ID}$, Seq, false>**, which will be used in request-response binding and protect to reply attack. A third binary value indicates a state of reply.

When an intermediate node, according to example node **A**, receives a **REQ** message at first time, inserts it into its reverse table **RT$_S$** and broadcasts further. Every next packet identified with the pair **<REQ, Seq>** is simply discarded. This process continues with every node in the network including destination node **D**. Node **D** broadcast message further in order to conceal fact that it is a real request destiny. This countermeasure compensate fact that **R$_{ID}$** is a hash value on destination pseudonym. The computation overhead of request retransmission for the entire network is negligible since every node does not relay the same requests twice. Note, that this move increase additional power consumption by **D**, what is a painful price for being untraceable. Similar approach we can find in [3]. Due to performance and memory consumption every entry in **RT$_S$** table should be removed after some idle time determined by **TIMER**.

**R$_{ID}$** is introduced to avoid computation overhead of public-key **PK** cryptography on every intermediate node. This improvement allow to use linearly scalable **PK** mechanism without expensive so called 'trapdoor' effect, where every node tries to decrypt received message checking whether it is legitimate destination.

| Step | Sender | Action | Receiver |
|---|---|---|---|
| 1 | S: M$_S$ = S$_{ID}$, ISeq, L$_P$, P | | |
| 2 | S: R$_{ID}$ = h(D$_{ID}$) | | |
| 3 | S: REQ, R$_{ID}$, Seq, E$_{PK\_D}$( M$_S$, Sig$_S$(M$_S$) ) | → | A: REQ, R$_{ID}$, Seq, E$_{PK\_D}$( M$_S$, Sig$_S$(M$_S$) ) |
| 4 | S: RqT$_S$ ← < R$_{ID}$, Seq, false > | | A: If < R$_{ID}$, Seq, S > is inside RT$_A$ do nothing, else continue |
| 5 | | | A: RT$_A$ ← < R$_{ID}$, Seq, S > |
| 6 | | | A: If h( A$_{ID}$ ) ≠ R$_{ID}$ then go to step **3** and broadcast **REQ** message, else go to **Phase II** |
| 3-6 | **A:** step **3** | → | **B:** steps **3-6** |
| 3-6 | **B:** step **3** | → | **C:** steps **3-6** |
| 3-6 | **C:** step **3** | → | **D:** steps **3-6** |

**Figure 3. Initialization/request of anonymous authentication – Phase I**

## 3.2 Anonymous reply – Phase II

When **REQ** message comes to destination, in example **D**, this node should be able to find out received **R$_{ID}$** on the hash values list of all using pseudonyms, Figure 4. It is allowed to use many unique pseudonyms for one identity and if it is applicable in reply message node **D** may suggest to source node **S** which one is the most preferable for future use, denoted as optional parameter **D$_{ID}$***. It makes stronger mechanism of protection against to traffic analysis attacks. In the next step **D** decrypts received message and verify its signature, then continues if succeeds. For request message node **D** can find in its reverse table **RT$_D$** an intermediate node to send a response, in example node **C**. Since the reverse path is well defined the response message may be sent to the source node **S**. Before that, an additional layer of anonymous symmetric cryptography is introduced to conceal the real content of response message. The node **D** generates a session key **K$_N$ = K$_{DC}$** for an anonymous communication with first intermediate node along reverse path direction. Then transports this key to chosen node, e.g. **C**, using secure transport protocol such as Elgamal or Shamir's no key protocol [25]. In the next step the node **D** creates its message **M$_D$** as it is illustrated in Figure 4. The purpose of added to **M$_D$** a session key **K$_{SES\_D}$** is to create the secured symmetric-cryptography based tunnel between nodes **D** and **S**. In this tunnel acknowledge message **ACK** and further communication will be performed. Since Phase II every type of message, in this case **REP**, is hidden in a part of **M$_{D/S}$**. Then created message **M$_D$** is encrypted with **PK$_S$** of source's node. Message format **M$_C$** destined to the **C** contains pair of **<R$_{ID}$, Seq>** used to follow the reverse path routing and additional **K$_{NN}$ =K$_{CB}$** known by **D, C** and in the next step **B**. This approach is intended for a reputation system purpose. Whenever node **C** sends data to **B** in behave of **D** the former will be able to evaluate whether data was forwarded correctly or if was sent further at all. These features impose on monitoring nodes an additional computation overhead and may be optional. Note, that this feature may be used on every willing node or only source and destination excluding intermediate ones as those who do not participate and are not interested in the quality of forwarding service. However it is highly encouraged to monitor every activity in the ad hoc network due to the reputation system ability of malicious nodes detection, which will be presented in later subsection. Note, that described approach does not reveal any information for passive attacker. Keys **K$_{NN}$ = K$_{SN}$* = K$_{DN}$*** are simple ignored by source and destination nodes.

In the next step the previously prepared and encrypted message **M$_C$** is sent to the node **C**. At the same time **D** updates its request and forward tables as it is depicted in Figure 3. Additional random identifier **F$_{ID}$** is issued to allow independently forward packet messages in both directions. This step is required if both authenticated sides are mobile along already established path. More detailed description of this feature presents next subsection. The intermediate node **C** decrypts the message $M_C$ and checks out for $< R_{ID}, Seq >$ in its request table $RqT_C$ whether it is a response to its request. If all rows in the table does not match aforementioned pair of $< R_{ID}, Seq >$, then it looks for next hop node in $RT_C$, namely **B**, and updates appropriately its forward table $FT_C$. Next, the node **C** establishes the secure connection with node **B** in order to forward the reply message further. The procedure continues cyclically for nodes **C,B,A** until the destination node **S** receives a response to its request.

| Step | Sender | Action | Receiver |
|---|---|---|---|
| 1 | D: $h(D_{ID}) == R_{ID}$ | | |
| 2 | D: $M_S = S_{ID}, ISeq, L_P, P, Sig(M_S)$ | | |
| 3 | D: Verify $Sig(M_S)$ | | |
| 4 | D: choose from $RT_D$ node for $R_{ID}, Seq => C$ | | |
| 5 | D: $SecK(K_N), K_N = K_{DC}$ | ↔ | C: $SecK(K_N)$ |
| 6 | D: $M_D = REP, F_{ID}, D_{ID}*, K_{SES\_D}, ISeq+1, L_P, P$ | | |
| 7 | D: generate $K_{NN}$ for next hop – B, | | |
| 8 | D: $M_C = R_{ID}, Seq, F_{ID}, K_{NN}, E_{PK\_S}(M_D, Sig_D(M_D))$ | | |
| 9 | D: $E_{K\_N}(M_C)$ | → | C: $E_{K\_N}(M_C)$ |
| 10 | D: $RqT_D \leftarrow <R_{ID}, Seq, true>$ <br> D: $FT \leftarrow <R_{ID}, Seq, D, null, C, K_N>$ <br> D: $FT \leftarrow <F_{ID}, Seq, C, K_N, D, null>$ | | C: $M_C = D_{K\_N}(M_C)$ |
| 11 | | | C: If $< R_{ID}, Seq >$ is inside $RqT_C$ go to **Phase III**, else continue |
| 12 | | | C: choose from $RT_C$ node for $R_{ID}, Seq => B$ |
| 13 | | | C: $FT \leftarrow <R_{ID}, Seq, D, K_{DC}, B, K_{CB}>$, <br> C: $FT \leftarrow <F_{ID}, Seq, B, null, D, null>$ <br> $K_{DC} = K_N, K_{CB} = K_{NN}$ |
| 5-13 | C: steps **5,7-13** for $K_N = K_{CB}, K_{NN} = K_{BA}$ | ↔ | B: steps **5, 9-13**, for $K_N = K_{CB}, K_{NN} = K_{BA}$ |
| 5-13 | B: steps **5,7-13** for $K_N = K_{BA}, K_{NN} = K_{AS}$ | ↔ | A: steps **5,9-13**, for $K_N = K_{BA}, K_{NN} = K_{AS}$ |
| 5-13 | A: steps **5,7-13** for $K_N = K_{AS}, K_{NN} = K_{SN}*$ | ↔ | S: steps **5,9-13**, for $K_N = K_{AS}, K_{NN} = K_{SN}*$ <br> S: $FT \leftarrow <R_{ID}, Seq, A, K_{AS}, S, K_{SN}>$ <br> S: $FT \leftarrow <F_{ID}, Seq, S, null, A, null>$ <br> $K_{SN}*$ is ignored |

**Figure 4. Anonymous reply – Phase II**

### *3.3 Anonymous authentication – Phase III*

In the last phase the titled authentication is performed and secure channel between source and destination nodes is established. The response to request of **S** may come from several different locations. It depends on how many responses were sent by **D** using different nodes. However, **D** may select the most trusted intermediate nodes in its nearest neighborhood but it is impossible to manage by **D** the routing tables along end–to–end path. It means that a set of paths may cross each other increasing computational requirements. So, it is advised to support only non-crossing paths. It may be achieved checking entries inside a forward table. Precisely, if the forward table has an empty entry matching $<R_{ID}, Seq>$ the path may be established.

The Phase III starts when the source node, say **S**, receives reply messages, what is shown in Figure 5. This happen at first time when the given message is identified by $< R_{ID}, Seq >$ and in **S**'s request table $RqT_S$ the reply status is false. Otherwise message is dropped end internal reputation system is informed about this incident.

In the next step a source node decrypts the received message and verify validity digital signature. If the process finishes successfully the node **D** is authenticated with the source node **S**. Then **S** is ready to prepare confirmation message **ACK**, optionally inserting to it preferable pseudonym for future use $S_{ID}*$. A created acknowledge is encrypted with proposed by **D** session key $K_{SES\_D}$ and then sent to the intermediate node using its secure tunnel. Sessions keys between forwarding nodes $K_N$ are unidirectional ($K_{AB} \neq K_{BA}$), so relayed **ACK** data are preceded by key exchanging phase. This approach provide a flexible mechanism for reputation monitoring and its using depends on every node power capabilities, in particular the only source and destination nodes may support this feature.

The **ACK** message traverses through the network form the source to the destination, using unicasted and secured channel. If the destination node is able to successfully decrypt and verify **ISeq** number this means that the source node **S** has been already authenticated with **D**.

| Step | Sender | Action | Receiver |
|------|--------|--------|----------|
| 1 | **S:** < $R_{ID}$, Seq > is inside $RqT_C$ | | |
| 2 | **S:** check status of **Reply**: if **Reply == false** then **Reply = true** and continue else **STOP** | | |
| 3 | **S:** from step 9 in Phase II $M_S = R_{ID}$, Seq, $F_{ID}$, $K_{NN}$, $E_{PK\_S}( M_D, Sig_D(M_D) )$ | | |
| 5 | **S:**$M_D$=REP, $D_{ID}$*, $K_{SES\_D}$,ISeq+1, $L_P$, P,$Sig_D(M_D)$ | | |
| 5 | **S:** Verify **Sig** ($M_D$) | | |
| 6 | **S:** $M_S$ = ACK, $S_{ID}$*, ISeq+2, $L_P$, P, | | |
| 7 | **S:** choose from $FT_S$ node for $R_{ID}$, Seq, to S => A | | |
| 8 | **S:** SecK( $K_N$ ), $K_N = K_{SA}$ | ↔ | **A:** SecK( $K_N$ ), $K_N = K_{SA}$ |
| 9 | **S:** $M_A = E_{K\_SES\_D} ( M_S )$ | | |
| 10 | **S:** $E_{K\_N} ( F_{ID}$, Seq, $K_{NN}, M_A ), K_{NN} =K_{AB}$ | → | **A:** $E_{K\_N} ( F_{ID}$, Seq, $K_{NN}, M_A ), K_{NN} =K_{AB}$ |
| 11 | | | **A:** choose from $FT_A$ node for <$R_{ID}$, Seq, S> => B and update $FT_A$ < $F_{ID}$, Seq, S, $K_{SA}$, B, $K_{AB}$ > $FT_A$ < $R_{ID}$, Seq, B, $K_{BA}$, S, $K_{AS}$ > $K_{SA} = K_N, K_{AB} = K_{NN}$ |
| 12 | | | **A:** If < $R_{ID}$, Seq > is inside $RqT_A$ update **Reply** status, decrypt content and verify **ISeq+2,** else continue |
| 7-12 | **A:** steps **7-10**, for $K_N = K_{SA}$ , $K_{NN} =K_{AB}$ | ↔ | **B:** steps **8-12**, for $K_N = K_{SA}$ , $K_{NN} =K_{AB}$ |
| 7-12 | **B:** steps **7-10**, for $K_N = K_{AB}$ , $K_{NN} =K_{BC}$ | ↔ | **C:** steps **8-12**, for $K_N = K_{AB}$ , $K_{NN} =K_{BC}$ |
| 7-12 | **C:** steps **7-10**, for $K_N = K_{BC}$ , $K_{NN} =K_{CD}$ | ↔ | **D:** steps **8-12**, for $K_N = K_{BC}$ , $K_{NN} =K_{CD}$ |

**Figure 5. Anonymous acknowledge, end of authentication process – Phase III**

In order to allow communication in both directions additional relation between $R_{ID}$ and $F_{ID}$ must be stored in a source and destination nodes. In real implementation during keys $K_N$ exchanging an extension to presented protocol is required. Every node should support a table binding values **<Originating node ID, $K_{N\_ID}$, $K_N$>**, where $K_{N\_ID}$ should be attached to the plain text header of every sent message. This key identifier should be changed with every message to disallow data tracing. This extension does not break the anonymity and requires a small computation overhead to every node.

## 3.4 Anonymous path maintenance

Already authenticated nodes can anonymously exchange data between each other. In the mobile environment these nodes usually move. The presented protocol supports mobile nodes in the following way. If source or destination node change its location loosing a radio range to the first intermediate node then it has to generate a request message with valid <$R_{ID}$,Seq> or <$F_{ID}$,Seq> pairs respectively for **S** and **D** node. This message is broadcasted to the entire network and reaches the second party of the couple <S,D> or one of already engaged intermediate nodes. In the first case the replay to the request originator is sent establishing a new anonymous path. If the second case happens, an intermediate node generates a reply to the source of request. In these scenarios all steps regarding communication between intermediate nodes must be applied. Instead of encrypted authentication payload intermediate nodes should insert padding. Appropriate reply message format is as follow: **<REP, $R_{ID}/F_{ID}$, Seq, $K_{XY}$*, $L_P$,P >**

In the case of unexpected break down of the anonymous path (e.g. power supply down in one of the nodes, node's movement) the intermediate node if it is possible should generate in reverse direction undelivered message error in format **<ERR, $R_{ID}/F_{ID}$, Seq, $K_{XY}$*, $L_P$,P >**. After some time determined by **TIMER** unused forward table entries related to < $R_{ID}/F_{ID}$, Seq > should be removed.

## 3.5 Reputation system

In an open environment such as mobile ad hoc networks anonymous interactions with unknown entities may be a source of behaviors independent from expectations. Trust management based on the reputation system is essential part facilitating a prediction and improving a performance of anonymous communication. In this paper we propose the distributed reputation system which is adaptation of Liu [14] reputation model.

Our reputation depends on time, own past experience, second-hand information and is expressed by level of trust. This input features are organized with reputation dynamic evaluation scheme providing nodes assessment. Proposed model consist of:

- $SR_B(A)^t$ – service reputation held by **B** expressing level of trust to node **A** in time **t**, and is taken into account whenever **B** is going to interact with **A**.
- $IR_B(A)^t$ – information reputation held by **B** expressing level of trust for information received from node **A** in time **t**, and is used for evaluating received second-hand information form **B**.
- $V_B(A)$ – second-hand information (vote) containing a recommendation of trust to node **A**, for an honest node is equal $SR_B(A)^t$.
- $ST_B(A)$ – satisfaction degree of node **B** during interaction with the node **A**.
- $OE_B(A)^t$ – own experience of node **B** based on history of interactions with the node **A**.
- Nodes exchange **V** information with truthful neighbors.
- Aforementioned parameters varies in range [-1,1], where the most positive value reflects to the most trustworthy parameter.

Every own experience of node **B** at time $T^{NEW}$ based on history of simple interaction with the node **A** is updated as follows [14]:

$$OE_B(A)^{T\_NEW} = OE_B(A)^{T\_OLD} * \rho^{T\_NEW - T\_OLD} + ST_B(A) * (1 - \rho^{T\_NEW - T\_OLD}),$$

where fading factor $\rho = [0..1]$ and determine influence history and new satisfaction degree to the own experience.

Whenever a new first **ST** or second hand information **V** is obtained the service reputation **SR** is updated. **SR** consists of own experience and weighted average of votes taking into account the information reputation **IR** of recommending nodes. Considering group **G** of voting nodes on **A**, the node **B** takes only those whose **IR** is positive. Own information is usually more valuable [16], [22], hence scaling factor $\alpha = [0..1]$ is introduced to the formula:

$$SR_B(A)^{T\_NEW} = \alpha * OE_B(A)^{T\_NEW} + (1 - \alpha) * \Sigma_P ( IR_P(A)^{T\_NEW} * V_P(A) ) / \Sigma_P IR_P(A)^{T\_NEW}$$

Note that nodes cooperating rarely have small service reputation **SR** and are not trustworthy in the case of new session establishing.

In order to evaluate credibility of recommendation **V** obtained from neighboring nodes it is required to update the information reputation **IR**. In this paper we propose a formula which considers close relations between node's experiences **OE** with particular node, say **A**, as well as other voter's **IR** and history of this information reputation.

$$IR_B(A)^{T\_NEW} = \beta * OE_B(A)^{T\_NEW} + \gamma^{T\_NEW - T\_OLD} * (1 - \beta) * \Sigma_P ( IR_P(A)^{T\_NEW} * (V_P(A) - OE_B(A)^{T\_NEW}) ) / \Sigma_P IR_P(A)^{T\_NEW},$$

where $\beta$ is scaling factor for own experience and $\gamma$ is a time fading factor, both lies in **[0..1]**. It is recommended to let the service reputation **SR** evolving more dynamically then information reputation **IR**, which means $\gamma < \alpha$.
The **IR** parameter depends on difference of our expectations and voter's information. This formula is different than the one proposed in [14]; it prevents to discredit a node with high service reputation by nodes being in collusion.

Every node during messages exchanging updates service reputation **SR** depending on its satisfaction level. The following list of behavior classes can be identified with anonymous authentication protocol we proposed:

- **Forwarding** – during network operations nodes are able to verify an integrity of messages anonymously forwarded in behave of them by overhearing the first intermediate node. Every message tampering, delays, double relays, dropping are detected as a malicious behavior,

- **Receiving** – every obtained message that could not be successfully verified, repeated messages with the same **ISeq,** break down a path without error message **ERR** with detectable presence of misbehaving intermediate node are treated as untrustworthy and should be take into account,
- **Anonymous path establishing** – an anonymous path establishing are three-pass process and in every phase multilayered operations are performed. By default every request packet **REQ** should be forwarder only once by every node. In the case of detection a behavior inconsistent with this rules or obtaining multiple copies of reply **REP,** error **ERR** messages reputation system should be informed.
- **Recommendation exchanging** – sharing reputation between nodes allow to compare own experience with a given by recommending nodes. In the case when the one of the votes differs much from the rest voters there exists presumption of node discrediting. Additional statistical cross-validation methods may be used for this case evaluation.

Interaction of system with protocol is performed ensuring purely anonymous communication. Information between nodes is sent on-demand of interested node. Every node may have temporal pair of public and private keys different from already assigned to nodes' pseudonyms. Request for recommendation consist of header with message type **REQV,** sequence number **Seq** and public key **TPK.** Every neighboring node should response to that message with a **TPK** encrypted list of pairs <node ID, V> regarding to its nearest neighborhood. Message completed with a header of **REPV** and hash value of (**Seq+1**) should be directly sent to the originator. A node initiating request message should save in its temporary memory the pairs of **<REQV, Seq>** and remove after some short time determined by **TIMER**. The responses are accepted only once from one node for each **<REQV, Seq>,** and every abuse of this rule should be notify to the reputation system. Described challenge-response scheme with exemplary nodes illustrates Figure 6.

| Step | Sender | Action | Receiver |
|---|---|---|---|
| 1 | A: REQV, Seq, TPK | → | B: REQV, Seq, TPK |
| 2 | | → | B: REQV, Seq, TPK |
| 3 | B: REPV, h(Seq+1), $E_{TPK}$(<A,V><C,V><Z,V>) | → | A: REPV, h(Seq+1), $E_{TPK}$(<A,V><C,V><Z,V>) |
| 5 | C: REPV, h(Seq+1), $E_{TPK}$(<A,V><B,V><Y,V>) | → | A: REPV, h(Seq+1), $E_{TPK}$(<A,V><B,V><Y,V>) |
| 5 | | | A:Verify h(Seg+1), $D_{TPK}$(<A,V><C,V><Y,V>) |
| 6 | | | A:Verify h(Seg+1), $D_{TPK}$(<A,V><B,V><Y,V>) |

**Figure 6. Anonymous second-hand information sharing**

## 4 Analysis of security against selected attacks

In this section the proposed protocol was exposed to analysis against know attacks taking into account an anonymity [6], [23], a security and reputation [14]. In this kind of analysis the two types of attacker model are assumed: passive and active. Passive attacker usually takes an advantage of overhearing capabilities whereas active attacker tries to interact with object of its wicked intentions. In the following analysis we assume attacker of both types of adversaries already gained an access to the network form trusted authority TA.

An anonymity is guaranteed when the sender, receiver and its relationship are undisclosed during all time of communication. The following types of the hostile activity are addressed to break down anonymity, reputation system and overall mobile ad hoc network:
- **Traffic and data analysis by passive attacker** – our protocol incorporates public and symmetric key cryptography with randomized tokens inside of each message. In the first phase of anonymous authentication only a source location and destination identifier may be known. In the second, third and during all the rest of communication only specialized methods of isolating all nodes along the path may reveal relation but only on the link layer (MAC). Neither node's pseudonyms nor its really identities be disclosed. Note, that in mobile environment this type of attack is difficult to conduct in along all multiple paths.
- **Message coding attack** – in this attack the adversary is trying eavesdropping the sending messages and to trace those in which the content matches the identified pattern. In our case the randomizing tokens based on padding covered by layers of symmetric encryption against to this attack.

- **Message length attack** – overhearing attacker tries to correlate all messages with the same length along chosen part or in the all mobile network. The padding we are using allow every node to setup the message length to the same constant value, what makes this attack ineffective.
- **Message replay attack** – an adversary tries to reply old packets repeatedly in order to identify a pattern of the same behavior during forwarding packets. During anonymous path establishing every doubled request messages are silently rejected, however messages repeated through the existing anonymous path will be delivered to legitimate receiver and dropped due to improper **ISeq** number. This would allow adversary to perform Denial of Service attack but the reputation system properly running quickly detects this misbehavior and isolate the hostile node. Additionally the cross-paths on every intermediate node may be allowed [3], [23], which decorrelates replied and forwarded packets. Finally we can enforce frequent changing symmetric keys between intermediate nodes, which completely change the content of the same repeated packets.
- **Collusion attack** (**active traffic analysis**) – similar to traffic analysis attack performed by an active adversaries being in collusion. In this attack we can assume the worst case when all intermediate nodes work together against the anonymity of source and destination. For this scenario the asymmetric and symmetric encryption in conjunction with random tokens (padding with the sequence number) performed between the source and the destination, effectively conceal a source's pseudonyms and all exchanged data. This type of attack in conjunction with message reply attack is able to reveal the locations of both communicating sides on the data link layer (MAC) but source anonymity is still ensured.
- **Timing attack** – attacker tries to correlate source and destination nodes taking an advantage of timing during communication. Times of start, stop, and duration of connection at sending node may be related with equivalent times of receiving data at destination node. This attack, similarly to former, may only point out the physical relation between nodes without revealing their identities. Additionally it effectiveness decreases when network load grows up.
- **Profiling attack –** is a long term method of tracing activity in the network. It is usually based on the message volume attack, counting the amount of sent data, and timing attack. Prevention and defense against to this attack is very hard. Adversaries creating a communication profile of sender and receiver may reveal pseudonym, even pseudonyms list of interested nodes. Even though secured channel hide vital data, where often changing session keys prevent from confidentiality compromise.
- **Sybil attack –** this attack take an advantage of using multiple identities by adversary's node. Multiple MAC addresses of data link layer cause a huge impact to the quality of services in ad hoc network because every time an attacker's node changes its identifier (Medium Access Control) it breaks down all already established paths. It is useful if abandoned identity achieve very low reputation in the network practically eliminating it from communication. An impersonation of new identity deceives reputation system. More sophisticated variety of Sybil attack assumes using multiple identities simultaneously what in conjunction with other types of attack may facilitate compromised nodes in cooperation and data exchange in collusion attack. The more advanced reputation system may monitor and correlate behaviors of several neighboring nodes [22], even those using multiple addresses as well. This improvement requires an additional power consumption and is not implemented in our proposal.
- **Man in the middle attack** – the proposed protocols is vulnerable to this attack in the case of session key exchanging phase between intermediate nodes. The protocol construction assumes end–to–end anonymous authentication whereas intermediate nodes become purely anonymous during all communication phases. However influence of this attack to an anonymity compromising is negligible because the end-to-end encryption layer hides all anonymous communication.

Mobile ad hoc networks, due to their openness suffer from several attacks directed to its routing and forwarding features [10], [20]. In our protocol proposal security mechanisms and reputation based neighborhood control makes it immune to many of them. The most dangerous attacks aim to denial of service. **Wormhole attack** is the one where defense against it is a big challenge for many known protocols in ad hoc networks. The wormhole attacker records all received messages, tunnels it to a different location and retransmit them. If all packets are tunneled there is no impact to the communication. But if the tunnel is able to forward, for instance **REQ** data, faster then other intermediate nodes and after that reject a reply message, then successfully breaks down communication. A defense to this attack was proposed in [13] and its main idea, called packet leashes take into consideration node parameter such as time, velocity, location and verify on the receiver side whether packet traverse the network in realistic time. Our protocol proposal is vulnerable to this attack, however wormhole method does not provide any mechanism for anonymity revealing.

## 5   Conclusions

In this paper we described a new proposal of anonymous authentication protocol for mobile ad hoc networks supplemented with enhanced distributed reputation system. Its main objective is to provide end-to-end anonymous and bidirectional path of communication in a dynamic and open environment such as MANET. The protocol delivers secure data links based on a public and symmetric key cryptography. The distributed reputation system monitors activity of nodes being along the path and evaluate their level of trust. In the complementary way it provides efficient and well protected mechanisms for the anonymity guarantees and defending to the known attacks.

The network initialization and maintenance are important features of MANET and determine its level of self-configuring and self-organizing capabilities. In the network setup phase our protocols and other approaches [3], [4], [8] as well take an advantage of trusted authority **TA**. The main goal of **TA** is key managing for all nodes participating in a network. We use **TA** for creating a revocable anonymity by means of users' pseudonyms. A big challenge for mobile at hoc network is to have an on-line truthful **TA** ensuring the secure and anonymous communication. Therefore the online identity and key management for MANET is an issue of further direction in our research.

A computational overhead and node's capabilities constrains is a big challenge for security design in MANET [10], [20]. Since the reputation system is promising approach for supporting anonymity we plan to incorporate methods with a less computational impact to every node. In terms of network performance we are going to perform simulations comparing it with existing approaches.